\documentclass[10pt]{iopart} 
\usepackage{iopams}
\usepackage{lipsum,babel}
\usepackage{graphics}
\usepackage{color}
\usepackage[usenames,dvipsnames]{xcolor}
\usepackage{dcolumn}
\usepackage{bm}
\usepackage[latin1]{inputenc}
\usepackage{textcomp}
\usepackage{footnote}
\usepackage{tablefootnote}
\usepackage{threeparttable}
\usepackage{hologo}
\usepackage{booktabs}
\usepackage{epstopdf}
\usepackage{url}
\usepackage{amssymb}
\usepackage{balance}
\usepackage{cite}
\usepackage{hyperref}
\usepackage{eqnarray}
\usepackage{graphicx}
\usepackage[labelfont=bf,font=small]{caption}
\def \MgB2 {MgB$_{2}$ }
\begin{document}

\title[Electromagnetic Assessment and AC Losses of Triaxial Cables]{Electromagnetic Assessment and AC Losses of Triaxial Cables with Multiple 2G-HTS Layers Per Phase}

\author{M. Clegg \& H. S. Ruiz}
\address{College of Science and Engineering \& Space Park Leicester, University of Leicester, Leicester LE1 7RH, United Kingdom}
\ead{mlc42@leicester.ac.uk; dr.harold.ruiz@leicester.ac.uk} 

\vspace{10pt}


\begin{abstract}
For an accurate estimation of the AC losses of superconducting triaxial cables, in this paper we present a two-dimensional model capable to provide a global assessment of multi-layer triaxial cables, validated against the reported AC-losses measurements on single-phase cables provided by the VNIIKP Cable Institute. Four models are presented, the first being a single-phase cable of 50 tapes and the others being three triaxial cables made of up to 135 coated conductors distributed in up to 9 layers. A systematic study is devised, where the number of layers per phase increases from 1 to 3, with at least 14 tapes distributed across each layer of the first (innermost) phase, 15 in the secondary (middle) phase, and 16 in the third (outermost) phase, respectively. Remarkably, our results reveal that the simple strategy of considering an unbalanced distribution for the amplitudes of the applied current, can generally balance the magnetic field between the three phases even for the bilayer and trilayer cables, resulting in negligible magnetic leaks in all situations. Besides, our high-resolution simulations allow to see for the first time how the transport and magnetization currents distribute across the thickness of all the superconducting tapes, from which we have found that the AC-losses of the 2nd phase is generally higher than at the other phases at low to moderate transport currents $I_{tr}<0.8I_{c}$. Nevertheless, depending on whether the $I_{c}$ of the SC tapes at the 3rd phase layers is lower than the one at the 2nd phase, the layers at the third phase can exhibit a considerable increment on the AC losses, this is as a result of the considered magneto angular anisotropy of the superconducting tapes, which lead to intriguing electromagnetic features that suggest a practical threshold for the applied transport current, being it $0.8I_{c}$. Likewise, the relative change in the AC-losses per adding layers, per phase, and as a function of the entire range of applied transport current is disclosed.
\end{abstract}

%
\ioptwocol
%
%
%


\section{Introduction}\label{Sec.1}

As current cabling technology continues to progress in line with the increased demands for medium and high voltage networks, the use of high-temperature superconducting (HTS) power cables offers the most flexible solution to the ever growing requirements of higher currents in reduced spaces, specially for projects that may involve highly restrictive right of way conditions. HTS power distribution cables have been developed since the early 2000s \cite{Fisher2003IEEE,Politano2001IEEE}, where manufacturing of both single and three-phased cables have advanced through the use of the so called first (1G) \cite{Demko2007IEEE,Stemmle2014IEEE} and second generation (2G) of HTS tapes \cite{Fetisov2017IEEE}. The natural evolution of cable designs has resulted in the search for increasing the transport current capabilities of these cables, whilst also reducing their effective cross-section, as only this can allow to scale down the size of the cryogenic system \cite{Malozemoff2015Book,Kalsi2011SC}, and reduce the costs of associated materials~\cite{Willen2005CIRED}. In this sense, a convenient solution to achieve both, the reduction in cost through the cryogenic system and, the increment on the current carrying capabilities of single or three-phase cables, is to simply utilize multiple layers of 2G-HTS tapes for each one of the phases. 

Nevertheless, just recently, only two research facilities have been capable to manufacture and test such cables at a real-scale level. In the first instance, a Korean team has developed a 23 kV/60 MVA class triaxial cable, which is to be installed in an undisclosed location of the Korea Electric Power Corporation (KEPCO) system~\cite{Lee2020Energies}. In this cable, only the innermost phase incorporates multiple layers of 2G-HTS tapes (two layers), whilst the other phases only have one. On the other hand, aiming to increase the transmission capacity of this triaxial cable, the Russian team at VNIIKP have taken a different route~\cite{Fetisov2021IEEE}, but using as a reference the Korean design. Within this alternative, which will become the focal point of this paper, rather than increasing the operating voltage of the cable, that implies to increase the insulation thickness and hence the cable diameter, their and our approach is to increase the operational current by using a multilayer structure for each one of the cable phases. The local (inside the superconducting tapes) distribution of current density, the magnetic field profiles, and the AC-loss predictions for the entire range of potential transport currents for this kind of cable, is presented in this paper.

Thus, in \autoref{Sec.2} a detailed description of all the cable designs considered in this study will be presented, it is followed by a brief summary of our computational models at  \autoref{Sec.3}. Then, the effect of augmenting the number of layers of 2G-HTS tapes per phase on local electromagnetic features such as the current density and magnetic field intensity across the cable, will be thoroughly discussed in \autoref{Sec.4}. Likewise, their relevant impact on the estimation of AC-losses for all the analyzed cables will be disclosed at \autoref{Sec.5}. Finally, in \autoref{Sec.6} the main conclusions of this study are summarized.


\section{Triaxial Cable Designs}\label{Sec.2}

Being interested in both, the manufacturing of real-scale power cables of single and three-phase current arrangements, and the designing of computational models capable to accurately predict the AC losses and relevant electromagnetic properties of these cables. The results of interest are to analyse the most recent advances on superconducting power transmission cables achieved by VNIIKP and SuperOx, where some experimental measurements of their critical current density and at some instances their AC-Losses, are already available for validation purposes. 


\begin{center}
\begin{table}[t]
\centering
\caption{\label{Tab:1} Parameters$^\dag$ of the single-phase trilayer plus bilayer-`shielded' cable in accordance with~\cite{Fetisov2018IEEE}.}
\begin{tabular}{ccccccccc}
\toprule
&$N_{l}$ & $R_{l}$ [mm] & $\Gamma_{l}$ [mm] & $N_{T}$ & $\langle I_{c0} \rangle$ [A]&\\
\midrule
&&&Cable Core&\\
\midrule
&1st&11.3&+200&8&72&\\
&2nd&12.2&+109&8&72&\\
&3rd&13&+61&8&72&\\
\midrule
&&&Cable `Shield'&\\
\midrule
&1st&18.4&-330&13&92&\\
&2nd&19.6&+110&13&92&\\
\bottomrule
\end{tabular}
\begin{tablenotes}
       \item $^\dag$ \footnotesize{$N_{l}$ stands for the number of layers, $R_{l}$ for inner diameter, $\Gamma_{l}$ for twist pitch length with the sign point at whether the winding direction is positive or negative, $N_{T}$ for number of tapes, and $<I_{c0}>$ corresponds to the averaged $I_{c}$ per tape per layer at self-field conditions but with the tape already wound into the triaxial arrangement.}
\end{tablenotes}
\end{table}
\end{center}

To this end, in this paper we will focus on a set of four different cables manufactured by VNIIKP, all of them making use of 4~$mm$ 2G-HTS SuperOx tapes, as related below:

\begin{itemize}
    \item[(\textit{i})] A single-phase power cable of 50 tapes in total (see \autoref{Tab:1}), 24 of them arranged in three layers of 2G-HTS tapes wound over the cable core, and the other 26 distributed in a further but relatively distant two layers with 13 tapes each. These outer two layers are commonly called by VNKIIP as the cable `shield'~\cite{Fetisov2018IEEE}, although their meaning must be understood within the right context of the experiments. Thus, the reader might find that the use of the word `shield' can be truly bewildering, as the experimental measurements to be reproduced are for having the same applied transport current to the `core-tapes' and to the `shield-tapes', indistinctly. This is because in the AC losses measurements reported by VNKIIP, both the `core-tapes' and the `shield-tapes' are connected in series, which means that we are actually dealing with a 5 layer single-phase cable. Still, the reader might wonder why the word `shield' is used by VNKIIP, and despite the above, why we decided to preserve this wording when referring to this specific cable design. The reason is twofold: Firstly, it is to be noted that the distance between the two sets of layers (cable core and cable `shield') is relatively large, which is for reducing the magnetization losses at the outer layers and; Secondly, the two outer superconducting layers in this cable may really act as a `magnetic shield' under adequate circumstances, such as by disconnecting them from the transport current source and by putting the cable under an external source of magnetic field, such as one created by a live neighboring cable. 
    \item[(\textit{ii})] The first Russian triaxial cable with a total of 87 tapes~\cite{Fetisov2017IEEE}, these arranged in a single-layer per phase distribution with the first phase containing 27 tapes, the second 29 tapes, and the third 31 tapes. 
    \item[(\textit{iii})] The most compact triaxial HTS power cable manufactured and tested up to date with currents up to 4 kA per phase~\cite{Fetisov2021IEEE}, it considering a double layered approach for each one of the phases, starting with 14 tapes per layer along the innermost phase, and increasing the number of tapes per layer for the second and third phases to 15 and 16 tapes, respectively (see \autoref{Tab:2}).
    \item[(\textit{iv})] And finally, an extension of the previous designed cable to consider a third superconducting layer per phase which is numerically realised.
    
\end{itemize}

\begin{center}
\begin{table}[t]
\centering\small
\caption{\label{Tab:2} Parameters for the multiple triaxial cables considered in this study$^{\ddag}$.}   
\begin{tabular}{ccccccc}
\toprule
Phase & $N_{l}$ & $R_{l}$ [mm] & $\Gamma_{l}$ [mm] & $N_{T}$ & $\langle I_{c0} \rangle$ [A]&\\
\midrule
1&1st&19.3&324&14&140.6&\\
&2nd&19.8&-171&14&147.9&\\
&3rd&20.3&125&14&147.9&\\
\midrule
2&4th&21.4&200&15&146.1&\\
&5th&21.8&-161&15&134.1&\\
&6th&22.2&131&15&134.1&\\
\midrule
3&7th&23.35&191&16&125.2&\\
&8th&23.75&-146&16&126.0&\\
&9th&24.15&123&16&126.0&\\
\bottomrule
\end{tabular}
\begin{tablenotes}
       \item $^\ddag$ \footnotesize{To see in conjunction with the single-phase cable definitions at \autoref{Tab:1} and, the properties of the bilayer triaxial cable prototype reported at~\cite{Fetisov2021IEEE}.}
\end{tablenotes}
\end{table}
\end{center}

Thus, it is to be noted that both of the aforementioned real-size prototypes of triaxial cables (\textit{ii} \& \textit{iii}) possess a similar number of tapes per phase, with 27, 29 and 31 in the single layered cable, whilst in the double layered structure, there are 28, 30 and 32 tapes evenly distributed across the two respective layers of each phase. Analogously, for the computationally devised cable (\textit{iv}), if the number of layers per phase increases to three where adding a further layer means an increment of just $\sim 0.5$~mm to the overall cable diameter, the number of tapes per layer at each phase is maintained. In other words, although changes on the overall diameter of the cable can be ignored within the context of these cables, the maximum transport current $I_{T}$ (below the sum of the minimum $N_{T}\{N_{l}\}\times\langle I_{c}\rangle$ per phase), increases in about $2$~kA per each added layer, i.e., in multiples of the $I_{T}$ of the single-layer triaxial cable (see \autoref{Tab:2}).

On the other hand, as the validation of any computational model for cables as complex as the ones described above depends strictly on the availability of experimental observations, and obviously, on the capacity of the model to reproduce such, geometry reductions from the real 3D cable topology to a simplified 2D-view of its transversal cross-section, can only be guaranteed if the averaged critical current density of the wound 2G-HTS tapes is directly measured by transport measurements on each of the 2G-HTS `layers' of the prototyped cable. In this way, $\langle I_{c} \rangle$ contains the three-dimensionality of the helically wound tapes which is not accounted in the classical consideration of the self-field critical current density of flat tapes, $I_{c0}$, commonly provided by the 2G-HTS manufacturer. This is where the success of our model lies and the reason why the VNIIKP cable prototypes are our study benchmark.

In particular, for the consideration of triaxial cables we have already demonstrated that as long as the $\langle I_{c} \rangle$ characteristics of the different cable layers are known~\cite{Clegg2022IOP}, as it was the case for the first triaxial HTS cables by VNIIKP~\cite{Fetisov2017IEEE}, and its successive developments~\cite{Fetisov2017IEEE,Fetisov2018IEEE,Fetisov2019,Fetisov2020,Fetisov2021IEEE}, it results possible to reduce the dimensionality of the problem to the classical 2D-scenario where the so-called H-formulation can be applied with ease. This is at least in what refers to implementing adequate constitutive physics within the computational model, i.e., including the magnetoanisotropic dependence $J_{c}(B,\theta)$ into the $\mathbf{E}-\mathbf{J}$ power law that governs the macroscopic electromagnetic properties of the superconductor~\cite{Ruiz2017SUST,Zhang2018}. This is quite advantageous in spite of having proven that 3D-models of similar cables, but with a much lower number of tapes (due to computing memory and processing requirements), could also be built within the same formulation and physical soundness~\cite{Fareed2022IOP,Fareed2022IEEE,Clegg2022IEEE}. However, for cables with tens or hundreds of 2G-HTS tapes as it is our case, the computing time and memory requirements simply result unaffordable. Still, worth mentioning is that not only for Y123-based tapes with a very high aspect ratio for the HTS layer $(~\sim 1\mu\mathrm{m}\times4\mathrm{mm})$, our 2D principles may be also applied to multilayered Bi2223 cables $(\sim0.24\mathrm{mm}\times3.8\mathrm{mm})$, where it has been recently demonstrated that the same 2D approach can be successfully applied for the estimation of the cable losses, achieving a difference with the experimentally measured AC-losses no greater than $10\%$~\cite{PetrovJoP2020}. 

\begin{figure}[t]
\centering
\resizebox{\columnwidth}{!}{\includegraphics{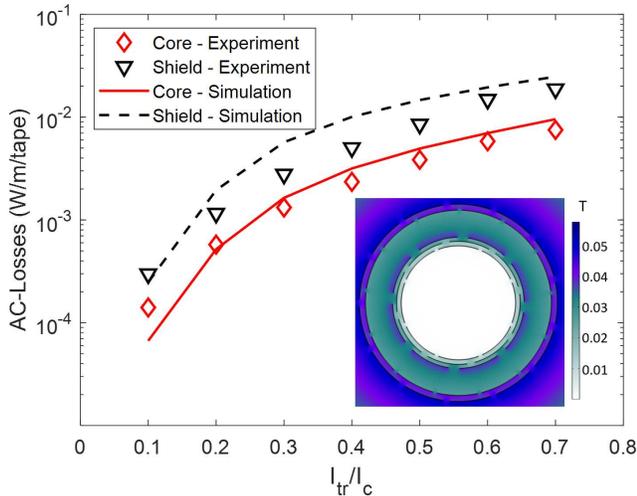}}
\caption{\label{Fig_1} Numerically calculated and experimentally measured AC-losses of the single-phase bilayer-shielded cable prototyped by VNIIKP~\cite{Fetisov2018IEEE}. In the inset, the profile of magnetic field intensity $|\mathbf{B}|$ for the cable at the first hysteresis peak, i.e., at $t=0.025~\mathrm{s}$ is shown. The applied transport current is $I_{a}=I_{tr}\sin(\omega t)$ with $I_{tr}=0.7 I_{c}$ and $\omega = 50~\mathrm{Hz}$.}
\end{figure}

Thus, due to all of the tapes featured in the considered cables being the $4$~mm wide tapes manufactured by SuperOx, consistency in modelling the tapes can be achieved through our prior knowledge of their physical properties and relative dimensions. This data can be found in Refs.~\cite{Lee2014IOP,Zhang2018,Fetisov2018IEEE,Clegg2022IOP}, with the main design parameters being summarized in \autoref{Tab:1} and \autoref{Tab:2}. Starting with cable design ($i$), i.e., the single-phase cable with fifty superconducting tapes distributed across 5 spaced layers (\autoref{Tab:1}), a few remarks are important to be stated. Firstly, as it can be seen from the bottom inset of \autoref{Fig_1}, where the computationally modelled cross-section of the cable is shown, for the time step being displayed the magnetic field outside of the cable is at its greatest value. It is due to having applied equal peak currents to the `core' and the `shield' layers, indistinctly. Under this experimental scenario, in the same figure the measured~\cite{Fetisov2018IEEE} and numerically calculated AC losses for the two sets of superconducting layers as grouped in \autoref{Tab:1} is shown, with a very good agreement in both cases, specially for the `core' layers. A slightly larger difference is seen for low to moderate currents, $i_{tr}=I_{tr}/I_{c}$, between 0.3 and 0.5, which might have been caused by having used a larger threshold for the critical current $\langle I_{c} \rangle$ at the outer (shield) layers, as reported in~\cite{Fetisov2018IEEE}, contrary to what have been seen in the most recent triaxial cables measured at VNIIKP\cite{Fetisov2021IEEE}. However, bearing in mind the inherent complexity of the cable and experimental conditions, as well as the relatively coarse mesh that has been assumed, i.e., where the superconducting tapes have been modelled by only three quad elements across their thickness, it is fair to state that the computational model has been duly checked. Thus, although this is not a triaxial cable as the one already studied in~\cite{Fetisov2017IEEE,Clegg2022IOP}, for the sake of this invitation, in this paper we decided to also present our results for the newest VNIIKP single-phase cable~\cite{Fetisov2018IEEE} as a benchmark model to validate our numerical results with their experimental measurements. 

\begin{figure}[!t]
\centering
\resizebox{\columnwidth}{!}{\includegraphics{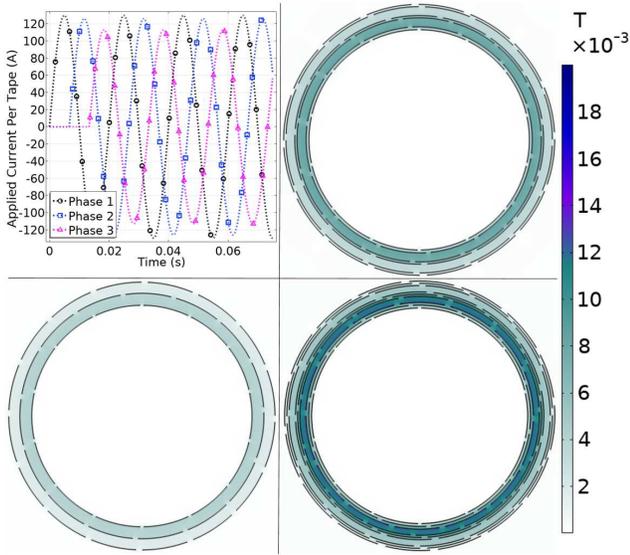}}
\caption{\label{Fig_2} The need for applying an unbalanced alternating transport current across the phases of triaxial cables is shown, it for balancing the magnetic field and avoiding magnetic leaks. At the top-left inset, the transport current strategy proposed by~\cite{Fetisov2021IEEE} is shown. It is followed by the 2D representation of the prototype cable at this reference in the right-top pane. Then at the bottom pane, the two additional triaxial cables considered in this study are shown, one being a monolayer triaxial cable at the left, and the other a trilayer triaxial cable at the right. The magnetic profiles correspond to the first hysteretic peak of phase 1, i.e., at $t=0.045$, with $I_{tr}/I_{c}=0.1$.
}
\end{figure}

The remaining cable designs all correspond to triaxial cables as the ones shown in~\autoref{Fig_2}. As for the prototype cable~\cite{Fetisov2021IEEE} (top-right inset), each phase is arranged into two layers with the number of tapes uniformly distributed across both of the layers. A total of 90 tapes are utilized with 28, 30 and 32 being the number of tapes for phase 1, 2 and 3 respectively, with phase 1 being the inner most layer, and with the main cable properties summarized in~\autoref{Tab:2}. Thus, for the method of investigation into the multiple layers of a triaxial cable, the outer layer of each phase is removed to create a single layered triaxial cable (bottom-left inset), with a total of 45 tapes dispersed as 14, 15 and 16 tapes per phase. Likewise, the final configuration being the addition of a third layer with a total of 135 tapes across the three phases with 42 tapes in phase 1, 45 in phase 2, and finally 48 in phase 3. The distance between the two layers in the real double layered triaxial cable is used as a reference, such that the same separation is used between the third and second layers of the trilayer triaxial cable. Likewise, the twist pitch has been coordinated so that the 2D representation is the same between the layers and finally the critical current for the third layer is assumed to be the same as the nearest layer, therefore giving a reasonable value to provide coherent results. An important point to notice here is that despite all the tapes being essentially the same, once wound into the triaxial arrangement, the actual critical current density of the 2G-HTS tapes at the different layers and phases can vary (see~\autoref{Tab:2}). Therefore, in order to balance the triaxial cable such as no magnetic leak is encountered outside of the outermost layer of the third phase, an unbalanced transport current is applied accordingly (top-left inset), which has been initially reported at~\cite{Fetisov2021IEEE}. 

\section{Numerical Modelling Strategy}\label{Sec.3}

By means of the general Partial Differential Equation (PDE) module of COMSOL Multiphysics, the single-phase and all the triaxial cables above discussed can be easily modelled within the so called 2D version of the H-formulation~\cite{Ruiz2019MDPI,Ruiz2019JAP,Clegg2022IOP}. However, considering the relatively large number of 2G-HTS tapes considered here, certain approximations need to be made in order to achieve reasonable computing times, whilst maintaining a sufficiently robust accuracy and sound physics. This means, to be able see the local electrodynamics of profiles of current density inside each one of the superconducting tapes, revealing the most characteristics elements of the vortex dynamics in the Bean's approach, such as the occurrence of magnetization currents countering the penetration of external magnetic fields, and likewise possible flux-free regions where neither the magnetic field nor the transport current have penetrated the entire superconducting domain at each one of the tapes. 

Thus, for being able to see the above in a 2D scenario, the $1~\mu$m thickness of the YBCO layer of the $4$~mm width 2G-HTS tapes at all our triaxial cables has been scaled up by a factor of 50. This allows to overcome the computational issues created by the aspect ratio of the superconducting tapes, but without significantly compromising the accuracy of the model. This strategy has been proven to be successful in many preceding studies~\cite{Stenvall2013SUST,Ruiz2019MDPI}, including here our own validation for the single-phase cable at \autoref{Fig_1} and, the monolayer triaxial cables studied at~\cite{Clegg2022IOP}. Therefore, the critical current density of the superconducting tapes at self-field conditions (\autoref{Tab:2}) is normalized accordingly, i.e., for each one of the phases/layers that compose the cable, including also their dependence with the angle and intensity of the magnetic field with respect to their main surface, following the $J_{c}(B,\theta)$ function 
%
\begin{eqnarray}
\label{Eq:JcBtheta}
J_{c}(B,\theta)=J_{c0}\left[1+\epsilon_{\theta}\left(\frac{B}{B_{0}}\right)^{\alpha}\right]^{-\beta} \, ,
\end{eqnarray}
which has been introduced at~\cite{Zhang2018} for the SuperOx tapes, with all the relevant parameters therein defined.

Each individual tape has a mapped mesh where the $1~\mu$m thickness of the 2G-HTS tape, now scaled to $50~\mu$m, has been split into 7 quad elements as previously shown in~\cite{Clegg2022IOP}, where a further discussion on the mesh quality can be found. The simulations are run for a total of 1.25 cycles of the transport current for the single-phase power cables and, 3.75 cycles for the three-phase cables. The latter is in order to ensure that all three phases are active at the time of choosing the peak-values for the hysteretic behaviour of the overall cable, with which the AC losses integral is to be resolved. 

\begin{figure*}[!t]
\centering
\resizebox{0.8\textwidth}{!}{\includegraphics{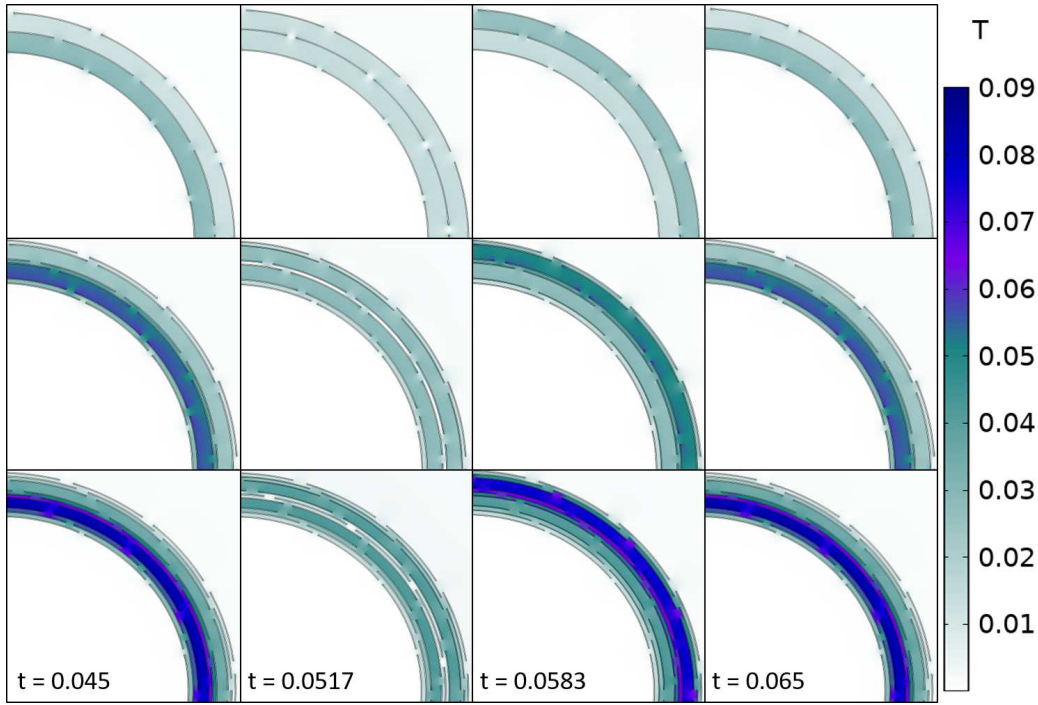}}
\caption{\label{Fig_3} Magnetic field profiles $|\mathbf{B}|$ within the hysteretic regime for the three triaxial cables considered in this study (top-to-bottom panes). These are being measured from the time step $t=0.045$~s (left pane) to $t=0.065$~s (right pane), i.e., from the third and forth positive peaks at phase 1 in~\autoref{Fig_2}, respectively. Two intermediate time steps are displayed, one corresponding to $t=0.0517$~s (Positive peak current at Phase 2) and $t=0.0583$~s (Positive peak current at Phase 3) for illustration of the magnetic field dynamics.}
\end{figure*}

Thus, although besides of the above mentioned the main features of the H-formulation and the 2D-modelling strategy for superconducting tapes could be considered as a classical subject, for the sake of completeness, in this paper we consider important to include its main functions within common jargon. In this sense, when employing the general PDE module of COMSOL within a 2-dimensional setting, the dependant variable, i.e., the magnetic field,  only has to be considered in the \textit{x} and the \textit{y} direction. Therefore, the current density flows perpendicular to the magnetic field, i.e., along the \textit{z}-axis, which is implemented through Faraday's law,
%
\begin{eqnarray}\label{Eq_Faraday}
\left| \partial_{y} E_{z}, -\partial_{x} E_{z} \right|^{\intercal}
=
-\mu \left| \partial_{t} H_{x} ,  \partial_{t} H_{y} \right|^{\intercal} \, ,
\end{eqnarray}
%
with Ampere's law being applied as the definition of the current density, 
%
\begin{eqnarray}\label{Eq_Ampere}
J_{z} = \partial_{x} H_{y} - \partial_{y} H_{x} \, .
\end{eqnarray}

Then, as is standard the electrical behaviour of the superconducting material is defined as the $E-J$ power law,
%
\begin{eqnarray}\label{Eq_EJ_law}
E_{z}=E_{0}\frac{J_{z}}{|J_{z}|}\left(\frac{|J_{z}|}{J_{c}}\right)^{n} \, ,
\end{eqnarray}
%
with the electric field criterion set as 1~$\mu$V/cm and, and the exponent $n=34.4$ according to the experimental measurements stated earlier for the critical current density dependence $J_{c}(B,\theta)$ for the $4$~mm SuperOx tapes~\cite{Zhang2018}.

On the other hand, regarding the hysteretic behaviour causing AC-losses within the 2G-HTS tapes, the calculation is done by integrating the local density of power dissipation $(\mathbf{E}\cdot\mathbf{J})$ across the said tapes. This is all completed over the time span for a full hysteretic cycle (\textit{f.c}), which as was stated earlier is why the simulations must be run for more than a single cycle. i.e., 
%
\begin{eqnarray}\label{Eq_Q}
Q = \omega \int_{f.c} dt \int_S \boldsymbol{E} \cdot \boldsymbol{J} dS \, .
\end{eqnarray}

Thus, making use of the above equations and experimental considerations for the magnetic field balancing of the triaxial cables, the H-formulation can be applied to reproduce numerically the experimental data, where the transport current is applied to each tape through integral pointwise constraints enclosing their cross-section.


\section{Electromagnetic Profiles}\label{Sec.4}

\begin{figure*}[!t]
\centering
\resizebox{0.8\textwidth}{!}{\includegraphics{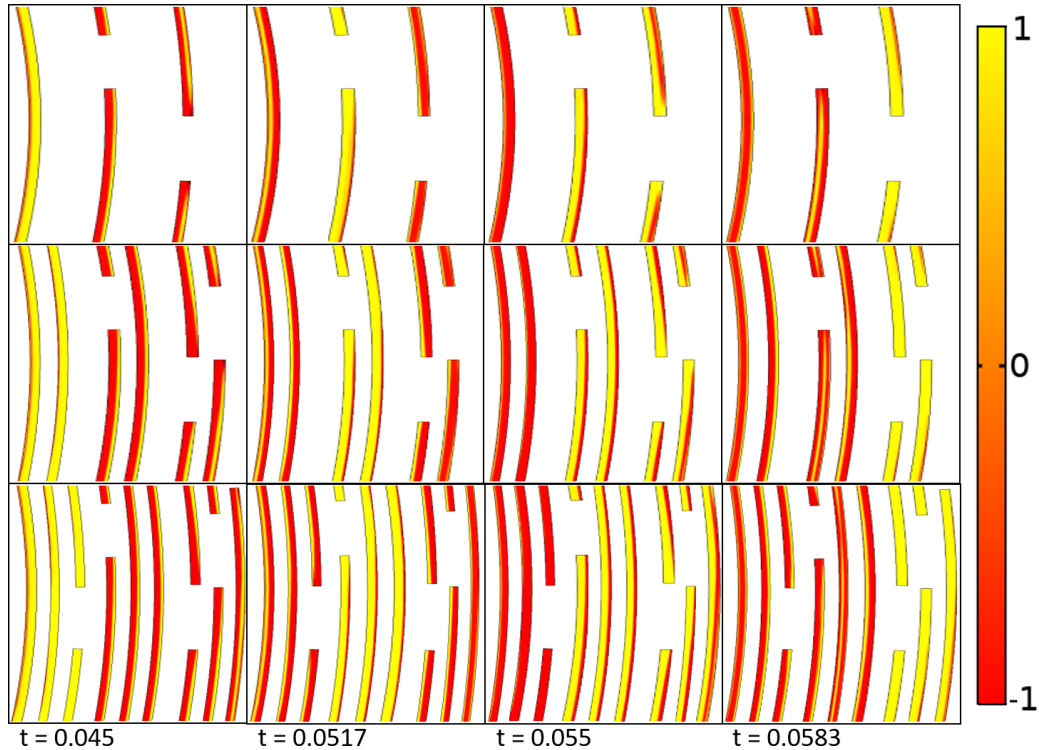}}
\caption{\label{Fig_4} Current distribution profiles $J_{z}/J_{c}$ in a sufficiently large section ($\sim\pm5^{\circ}$ over their x-axis) of (from top to bottom) the single, double, and trilayer triaxial cables shown at \autoref{Fig_2} and \autoref{Fig_3} where, the local distribution of currents inside the different superconducting tapes at the different phases and layers can be discerned. The profiles are shown for the time steps $t=0.045$~s and  $t=0.0517$~s at the two left panes, i.e, for the positive hysteretic peaks of $I_{tr}$ at phase 1 and phase 2, respectively (see \autoref{Fig_3}). Likewise, additional profiles for $t=0.055$~s and  $t=0.0583$~s are shown, these corresponding to the negative peak of phase 1 (third pane from the left), and the positive peak of phase 3 (right pane). A supplementary figure with a non fully compatible color scheme with B/W printing has been is provided, to allow a straight discerning of flux-free regions with $J_{z}=0$ from transport current and magnetization current regions with opposite $J_{c}$ current.}
\end{figure*}

To appreciate the interaction between the phases and tapes, analysis of the electromagnetic profiles is required to explain the differences in the AC losses. Particular interest has been paid to the magnetic fields at the peak transport current of 0.7 $I_{c}$ for each individual phase within the hysteretic cycle, which is being measured from $t=0.045$~s to $t=0.065$~s (see~\autoref{Fig_3}). This means that within the hysteretic cycle which commence after 2.25 cycles of having started with the applied current at phase 1 (according with \autoref{Fig_2}), the magnetic field profiles at the positive peaks for the applied current in phase 1 ($t=0.045$~s or $t=0.065$~s), phase 2 ($t=0.0517$~s), and phase 3 ($t=0.0583$~s) are displayed. Notice that the profiles displayed for the times $t=0.045$~s or $t=0.065$~s at the left and right panes of~\autoref{Fig_3} are identical, proving therefore the completeness of the hysteretic cycle. Also, it is worth mentioning that what is displayed in~\autoref{Fig_3} is the norm of the magnetic field density, which therefore means that exactly the same profiles of magnetic field could be observed for the negative peaks of current at each phase, hence providing a more comprehensive view of the hysteretic cycle. Thus, as it can be noticed in this figure, no magnetic field leaks can be seen for any of the triaxial cables considered, meaning then that the same transport current strategy introduced at \autoref{Fig_2} and originally proposed at~\cite{Fetisov2021IEEE}, has been proven to work regardless of the number of superconducting layers included per phase. Then, the most perceptible difference between the three triaxial cables from \autoref{Fig_3} is the increment of the magnetic field as the number of layers per phase increases, which is equivalent to the increment of the total transport current allowed by the cable. In finer detail, it is to be noticed that for the single layered cable (top-pane), and at the positive $1^{st}$ phase peak current (left pane), the intensity of the $\mathbf{B}$-field between the phase 1 and 2 layers doubles in average the field observed in between phase 2 and 3, with the maximum being up to 30~mT. Then, as the number of layers per phase doubles or triples for the other triaxial cables (middle and bottom panes, respectively), the maximum field increases to 60~mT and 90~mT, accordingly.

Likewise, at the positive peak of phase 2, the magnetic field is balanced at either side of the phase 2 layers in all of the cases, with the magnitude being 15~mT, 30~mT, and 45~mT as the number of layers per phase increases from 1 to 3, respectively. At the positive peak of phase 3, similar behaviour is observed to what was observed when phase 1 is at its positive peak, but now with the field intensity between phase 1 and 2 being approximately half of what is in between phase 2 and 3, with maximums again around 30~mT, 60~mT, and 90~mT, respectively. Thus, as the magnetic field intensity across the different layers of the triaxial cable vary within the range where significant changes in the critical current density of the SuperOx tapes have been observed~\cite{Zhang2018}, it results imperative not only to consider their corresponding $J_{c}(B,\theta)$ function (\autoref{Eq:JcBtheta}) in their computational modelling, but also to analyse the distribution of critical current density across each one of these superconducting tapes in the triaxial arrangements (see \autoref{Fig_4}), a matter that up to the best of our knowledge was uncharted, it due to the subjacent computational burden that implies such resolution.

For understanding the mutual interaction between the superconducting tapes at the different layers and phases of the triaxial cables, in~\autoref{Fig_4} we show the corresponding normalized current distribution profiles, $J_{z}/J_{c}$, for an applied transport current with a maximum amplitude of 0.7 times the lowest critical current measured at the real triaxial cable~\cite{Fetisov2021IEEE}. This means for the the third phase inner layer in the bilayer cable prototype, which corresponds to the seventh layer of the trilayer  triaxial cable accounted in \autoref{Tab:2}. Therefore, in agreement with~\cite{Fetisov2021IEEE}, the maximum transport current is applied to the innermost phase (phase 1) of the cable, as shown in the top-left inset of~\autoref{Fig_2}, with the other current-phases being adapted accordingly. Notice that this is necessary not only due to the need for balancing the magnetic field of the triaxial cable as explained above, but because the relative $\langle I_{c0} \rangle$ for the layers of superconducting tapes at the inner phase can be actually larger than the ones at the outermost phases. Consequently, this conditioning impose a physical threshold for the transport current on the entire cable, which furthermore must be applied to the innermost phase layers as a reference, as the superconducting tapes at the outermost phases can be seen, intuitively, more affected by the magnetic field produced by the tapes at the innermost phases.

Similarly, two more aspects need to be mentioned when analysing~\autoref{Fig_4}. The first being that having chosen a maximum transport current of $0.7 I_{c}$ when displaying the local profiles of current density is not a random decision, as it highlights the fact that between this and $0.9 I_{c}$, the cable is considered to be in proper conditions for its normal operation, i.e., under low fault risk scenarios. However, as higher is the amount of transport current introduced into a superconducting layer as less evidence will be seen of the magnetization currents, reason why assuming a peak current of $0.7 I_{c}$ seems the best alternative for untangling the current distribution profiles in realistic scenarios. On the other hand, slightly more cumbersome to understand is the impact of having each one of the superconducting layers at the triaxial cable with a different $J_{c0}$, and in fact a different $J_{c}$, the latter due to the magneto angular anisotropy of the superconducting tapes (see~\autoref{Eq:JcBtheta}). In this sense, as $J_{c}$ changes in time as $I_{tr}$ is a time-dependant sinusoidal function, to show a univocal representation of the critical current density across all the layers of the triaxial cable, $J_{z}$ must be normalized to the corresponding $J_{c}(B,\theta)$ at each one of the superconducting tapes, such that the general and sound physical principle of the critical state theory with $J_{z}\leq J_{c}$ is maintained, as it can be seen in~\autoref{Fig_4}.

Thus, for the monolayer triaxial cable with the profiles of current density shown at the top pane of \autoref{Fig_4}, the electrodynamics of profiles of current for each one the phases can be analysed with ease. This will help to understand the electrodynamics of multilayer triaxial cables, as at least in what regards to type-II superconductors, the physical principles involved are universal. For instance, notice that when the phase 1 is at its positive peak of applied current and within the hysteretic regime, i.e., for $I_{Ph1}=0.7I_{c}$ at $t=0.045 \mathrm{s}$ (see as reference the top-left pane of \autoref{Fig_2}), the transport current at the innermost layer of the 2G-HTS tapes composing such phase, flows mostly along their outer surface. However, as $I_{tr}<I_{c}$, a clear flux free region with absence of either magnetization or transport currents ($J_{z}=0$) can be observed near the innermost surface of the respective tapes, or in other words, at the internal face of the tapes wound over the cable former. Then, for the same instant of time, notice that the phase 2 of the applied current is at negative values but with positive slope. This means that in accordance with Bean's model, whilst most of the carrying current at the corresponding tapes (middle layer in the monolayer triaxial cable) is negative, $+J_{c}$ profiles must appear at the outer surface of largest radius (measured within the 2D perspective of the whole triaxial cable). A similar but more cumbersome behaviour is seen for the outermost layer of superconducting tapes, i.e., those connected to the phase 3 of $I_{tr}$, where the majority of the carrying current must be also negative, but where a closed flux front with $J_{z}=0$ is clearly discerned. This closed flux front is caused by the consumption of pre-existing magnetization currents with $J_{z}=+J_{c}$ at the outer surface of the superconducting tapes, it caused by the negative slope of the applied transport current, a phenomena that has been explained in great detail for the case of bulk wires subjected to synchronous and asynchronous AC conditions in~\cite{Ruiz2012APL,Ruiz2013IEEE,Ruiz2013JAP}. Then, from this knowledge, it has to be noticed that exactly the same phenomenological behavior can be observed from the local distribution of current density in the bilayer and trilayer cabes considered in the mid and bottom pane of \autoref{Fig_4}, respectively.

Then, as the time increases the dynamics of the current density profiles matches the critical-state physics above described, together with the corresponding magnitude and slope of the applied transport current for each one of the three phases. Therefore, just for the sake of completeness, the reader could easily identify the following from \autoref{Fig_4}: \textit{(i)} at $t=0.0517~\mathrm{s}$ (2nd column of plots from left to right) phase-2 is at its positive peak current and therefore the tapes at the middle layers display a similar if not almost identical distribution of current than what was observed for the phase 1 at $t=0.045~\mathrm{s}$. Therefore, the distributions of current across the superconducting tapes within the phases 1 and 3 at $t=0.0517~\mathrm{s}$ are comparable (qualitatively) with their peers at the phases 3 and 2 when $t=0.045~\mathrm{s}$, respectively. A similar behaviour occurs then when the phase-3 is at its positive peak, i.e., at $t=0.0583~\mathrm{s}$ where a similar analogy could be made between the current distributions of phase 1, phase 2, and phase 3, with the distributions of current observed in this order, at phase 2, phase 3, and phase 1 at $t=0.045~\mathrm{s}$, or at phase 3, phase 1 and phase 2 at  $t=0.0517~\mathrm{s}$, respectively. Then, \textit{(ii)} at $t=0.055~\mathrm{s}$ (3rd column of plots), i.e., when the transport current at phase-1 is at its negative peak, the local distribution of current profiles for each one of the phases must match the one attained when $t=0.045~\mathrm{s}$, but with their currents flowing in opposite direction. This means that any region with $\pm J_{c}$ at $t=0.045~\mathrm{s}$ is inverted to $\mp J_{c}$ at $t=0.055~\mathrm{s}$, proving that just a half cycle within the considered hysteretic regime could have been used for the calculation of the AC-losses if~\ref{Eq_Q} is multiplied by a factor 2, but still it was necessary to run the simulations until $t=0.0583~\mathrm{s}$ to fully ensure the hysteretic cycle was correctly established. Nevertheless, it is worth emphasizing that the comparison made above for different characteristic time events is purely quantitative, as indeed slight differences in the local distribution of current density profiles can be seen between these events. Nonetheless, as the energy losses of the system is calculated as the dynamic response of the superconducting material to the motion of the flux front profile, it is not possible to assert that the energy losses of multilayer triaxial cables obeys a linear relation with the number of superconducting layers per phase, even if each layer maintains the same number of 2G-HTS tapes considered within the monolayer case. Therefore, a careful inspection of the AC-losses of triaxial cables is due in the following section.


\section{AC-Losses}~\label{Sec.5}

\begin{figure}[t]
\centering
\resizebox{0.97\columnwidth}{!}{\includegraphics{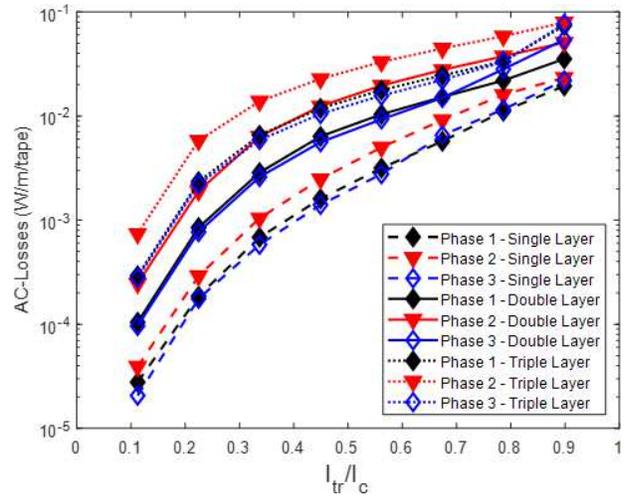}}
\caption{\label{Fig_5} The predicted AC-losses per tape of the simulated double layer triaxial cable compared with the losses produced by a single and triple layered triaxial cable.}
\end{figure}

\begin{figure}[!t]
\centering
\resizebox{1.0\columnwidth}{!}{\includegraphics{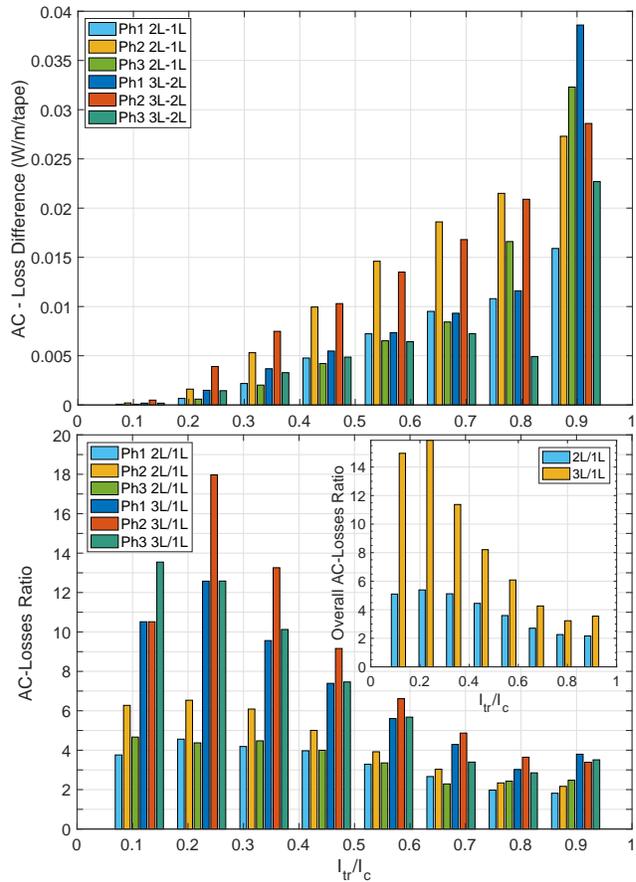}}
\caption{\label{Fig_6} Bar charts displaying the growing rate and ratio for the AC-losses of different triaxial cables as a function of the applied transport current $I_{tr}$ and their number of superconducting layers (L) per phase (Ph). Notice that for a better visualization, each of the eight set of six bars corresponds to the following $I_{tr}/I_{c}$ values 0.112, 0.225, 0.337, 0.449, 0.562, 0.674, 0.786, and 0.898, in the respective order.
}
\end{figure}

From the electromagnetic profiles, further analysis can be applied to the energy losses generated by the cables, where as explained in previous sections, a fully hysteretic behaviour has been found between the times $t=0.045$~s and $t=0.065$. Evidently, as the number of layers per phase augments the current carrying capabilities of the triaxial cable increases and consequently its AC-losses. However, the AC losses does not increase linearly with the augment of layers, i.e.,  of superconducting tapes, as it can be seen in~\autoref{Fig_5}. In other words, by doubling or tripling the number of layers of 2G-HTS tapes whilst aiming to maintain the overall dimensions of the cable, the AC-losses does not simply double or triple but actually increase substantially, specially for the low transport current regime (see also~\autoref{Fig_6}). 

Moreover, when inspecting in detail \autoref{Fig_5}, it is possible to see that regardless the number of layers considered into the triaxial cable, the AC-losses of the secondary phase is always greater than in the primary and tertiary phases. In fact, the AC-losses of the primary (innermost) and tertiary (outermost) phases is almost the same in all cases, at least for currents $I_{tr}\leq0.7I_{c}$. This is because, on the one hand, for the innermost phase magnetization currents are only produced by the superconducting tapes located within the same phase. This is due to the geometry of the cable, as by Faraday's law the magnetic field enclosed by the arrangement of $J_{z}$ profiles along the $xy-$plane is zero (neglecting the edge effect at each individual tape). Moreover, for the tertiary (outermost) phase, the magnetic field observed by such superconducting layers also is nearly zero, as its self-field has been compensated by the fields created by the $\pm120^{\circ}$ de-phased currents at the primary and secondary phases. On the other hand, the behavior of the secondary (middle) phase is different, as at this region the magnetic field created by the primary phase is not compensated by the tertiary phase as per Faraday's law, which agrees with our previous observation of a larger amount of magnetization currents occurring in the 2G-HTS layers at the middle of the cable (Phase 2). Then, in this order of ideas, for transport currents approaching the threshold value $I_{c}$, the AC-losses of all three phases render towards the same value, as this can be appreciated at $I_{tr}/I{c}=0.9$ in \autoref{Fig_5}. This is due to the phenomenon of the consumption of magnetization currents by the injected transport currents, and is previously explained for the case of isotropic superconducting monowires at~\cite{Ruiz2012APL,Ruiz2013IEEE,Ruiz2013JAP}. 

Nevertheless, for the case of the bilayer and the trilayer triaxial cables, the AC losses curves reveal that for transport currents as high as $I_{tr}=0.8I_{c}$, the magneto-angular anisotropy of the superconducting tapes~\cite{Ruiz2017SUST,Zhang2018} and their edge effect over its neighboring ones, imply the possibility to briefly encounter small regions (weak spots) over some of the superconducting tapes where the local $I_{c}$ becomes temporarily smaller than the $\langle I_{c0} \rangle$, this due to the time-dependence of the field $B$ caused by our AC regime. However, we must be clear when mentioning that at this stage it is not possible to state, for certain, whether the appearance of these naturally occurring weak spots, i.e., not formed by the inclusion of impurities nor defects into the superconducting tapes, are indeed metastable events of flux-pinning, or on the contrary these could lead to the actual destruction of the superconducting state by thermal quenching. This is because the simulations here presented are purely electromagnetic and do not consider Joule effects or changes in the temperature of the superconducting layers. Consequently, although a noticeable rise on the AC losses of the cable could be perceived at such high values of $I_{tr}$, the phenomena observed remain hysteretic. Nonetheless, in real life scenarios,  if the temperature of the overall cable cannot be maintained by an adequate flow and pressurizing of the LN2 coolant, the validness of \autoref{Eq:JcBtheta} cannot be assured. Consequently, based on a purely electromagnetic analysis the correct functioning of the triaxial cables can only be ensured or recommended for currents as high as $I_{tr}=0.7I_{c}$.

Also, a careful inspection of how the AC-losses change between the different triaxial configurations, i.e., the effect of adding further layers of superconducting tapes per phase, reveal quite intriguing features on the overall electrodynamics of the triaxial cables. This can be observed through~\autoref{Fig_6}, where the growing rate and ratio between the AC-losses per phase attained at the bilayer (2L) and monolayer (1L) cables, and the trilayer (3L) and bilayer cables are shown as a function of $I_{tr}$. Then, notice that the first three out of six bars at each $I_{tr}$, labelled from top to bottom at the figure's legend, represent the corresponding change in the AC losses per phase observed from the bilayer triaxial cable, this with respect the monolayer triaxial cable. Likewise, the remaining three bars display the relative difference (top plot) or ratio (bottom plot) between the trilayer and bilayer triaxial cables in the same sequence.

Thus, it is to be noticed that although for each cable design the relative change or increment in the magnitude of the AC losses is significant as $I_{tr}$ increases (see~\autoref{Fig_5}), when this is analysed with respect to the number of layers per se, the effect might be seen exacerbated when the relative difference between the AC losses for the 2L-1L or 3L-2L designs is depicted (see top pane of~\ref{Fig_6}). However, under a normalized scale, i.e, by calculating the ratios of AC losses between designs, i.e.,  2L/1L and 3L/1L, we can see that in reality it is quite the opposite (see bottom pane of~\ref{Fig_6}). For instance, notice that for low to moderate transport currents $(0.1 \leq I_{tr}/I_{c} \lesssim 0.6)$, the AC losses of the three phases of the bilayer cable at least triples the AC losses attained by the monolayer cable, with the losses of the secondary phase being up to 6 times for $I_{tr}/I_{c} \lesssim 0.3$, as it is the phase more prone to be affected by the magnetization currents as demonstrated in the previous section. In fact, we have found that the overall AC losses of the bilayer triaxial cable is about five times (in average) the ones of the monolayer cable for  $I_{tr}/I_{c} \lesssim 0.45$ (see inset at the bottom pane of~\ref{Fig_6}), and can double or even triple this as the current decreases when a further layer of superconducting tapes is added to each of the phases. Nevertheless, due to the already explained consumption of the magnetization currents by the transport current, the ratio of losses between the cables decreases as $I_{tr}$ increases, it tending to the physical threshold where the losses of the bilayer or trilayer cable are expected to double or triple the losses of the monolayer cable at $I_{tr}=I_{c}$. Still, it is worth emphasizing that for the most realistic regimes of $I_{tr}<I_{c}$, and indeed for the recommended working condition of $I_{tr}=0.7I_{c}$, i.e., where no metastable overcritical states are to be found within the hysteresis cycle, the impact of increasing the number of layers of 2G-HTS tapes in the AC losses of multilayer triaxial cables, is not a linear function of the losses for the corresponding monolayer cable. In fact, for the specific case of $I_{tr}=0.7I_{c}$, we have found that the losses of the bilayer and the trilayer triaxial cables are about 2.5 and 3.75 times the losses of the monolayer cable.


\section{Conclusion}~\label{Sec.6}

In this paper, our main aim has been to provide a comprehensive  assessment of the electromagnetic performance of recently proposed multilayer triaxial cables, where yet scarce experimental measurements are known~\cite{Fetisov2018IEEE,Fetisov2021IEEE}. This is due not only because of the great complexity of the cable designs, but in because in previous research it has been already made clear that for an adequate measuring of the total superconducting AC losses of the cable, including magnetization and transport current losses alike, an expensive and by-purpose large scale calorimetric rig must be built. This fact can effectively limit the actual deployment of projects aiming to implement triaxial cables in real power systems, reason why computational studies as the one presented here comes in a fore. 

Nevertheless, as electrical measurements of the critical current density of the 2G-HTS wires used for the fabrication of the triaxial cables are known,  both before and after the winding of the cable~\cite{Ruiz2017SUST,Zhang2018,Fetisov2018IEEE,Fetisov2021IEEE}, our numerical models have been demonstrated to offer realistic assessments for the AC losses of monolayer triaxial cables under single phase regimes, as well as bilayer triaxial cables under three-phase regimes. Then, based on these results, we have designed three different triaxial cables where each phase is defined by a monolayer, bilayer, or trilayer configuration with 14, 15, and 16 2G-HTS tapes wound per layer, it aimed to provide a systematic study on how the AC losses of these cables increase as a function of the transport current when doubling or tripling the number of superconducting tapes. This means the designing of superconducting cables with a total of 45, 90, and 135 tapes wound over a cylindrical former (see~\autoref{Tab:2}).

A full disclosure of the distribution of profiles of current density and magnetic field strength is provided for each one of the cables above mentioned, with sufficient resolution to see the occurrence of magnetization and transport current profiles within each one of the superconducting tapes. This allows to reveal the local dynamics of the so-called flux-front profile within the acclaimed critical state theory for type-II superconductors, which indeed governs the occurrence of hysteretic losses within the AC regime. In this sense, although we have demonstrated that the governing physics for the electromagnetic performance of triaxial cables continue to obey the general principles of the critical state theory, the magneto-angular anisotropy of the critical current density for the accounted superconducting tapes, can transitorily lead to what could be understood as metastable events of flux pinning at the phases most affected by the mutual inductive field, i.e., for those tapes wound at the secondary (middle) and tertiary (outermost) phase of the triaxial cable. This impose a certain limit on the maximum amount of transport current that from just the electromagnetic perspective, could be recommended for the operation of the triaxial cable, being this $0.7 I_{c}$, with $I_{c}$ the minimum averaged critical current density measured at the wound tapes in the triaxial arrangement.

Likewise, from a practical side some final remarks on the electromagnetic assessment of triaxial cables are to be mentioned, such as confirming that the experimentally measured and adopted strategy by VNIKKP of applying a three-phase current with unbalanced amplitudes, effectively serve to account for the differences in the critical current density of the wound tapes at the different layers/phases, leading to a correct balancing of the electromagnetic field, with negligible magnetic leaks at the outside of the cable. Moreover, we have proven that the relationship between the AC losses of triaxial cables with the number of possible layers of superconducting tapes per phase is not linear, and in fact can increase significantly at low to moderate transport currents $(I_{tr}/I_{c} \lesssim 0.45)$, with a minimum AC-losses of approximately 2.5 and 3.75 times the AC-losses of the monolayer cable at $I_{tr}=0.7I_{c}$ when the number of layers is doubled or tripled, respectively. 


\section*{Acknowledgements} This work was supported by the UK Research and Innovation, Engineering and Physical Sciences Research Council (EPSRC), grant Ref. EP/S025707/1, led by H.S.R. All authors acknowledge the use of the High Performance Computing Cluster Facilities (ALICE) at the University of Leicester (UoL). M.C thanks the CSE and EPSRC-DTP studentships provided by UoL. Networking support provided by the European Cooperation in Science and Technology, COST Action CA19108 (Hi-SCALE), is also acknowledged.

%
\section*{References}
\bibliographystyle{iopart-num}
\bibliography{References_Ruiz_Group}
%


\clearpage 

\begin{figure*}[p]
\centering
\resizebox{0.8\textwidth}{!}{\includegraphics{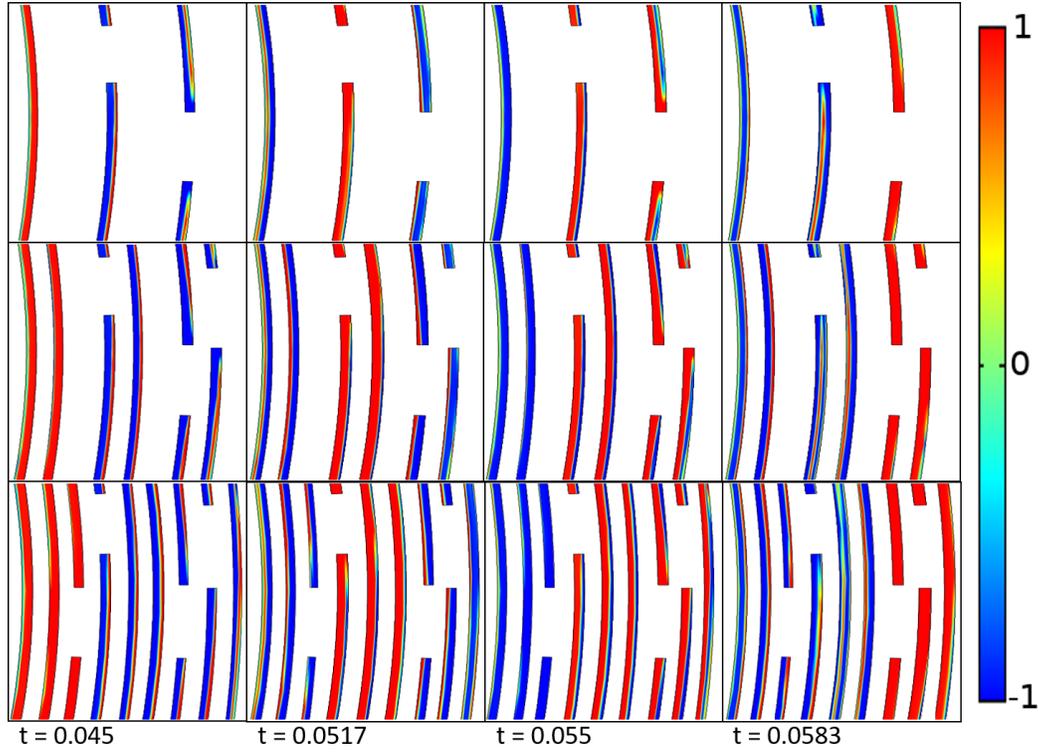}}
\caption{\label{Fig_4supp} \textbf{Supplementary figure.} Same as in Fig.~4 of the original manuscript. Current distribution profiles $J_{z}/J_{c}$ in a sufficiently large section ($\sim\pm5^{\circ}$ over their x-axis) of (from top to bottom) the single, double, and trilayer triaxial cables shown at \autoref{Fig_2} and \autoref{Fig_3} where, the local distribution of currents inside the different superconducting tapes at the different phases and layers can be discerned. The profiles are shown for the time steps $t=0.045$~s and  $t=0.0517$~s at the two left panes, i.e, for the positive hysteretic peaks of $I_{tr}$ at phase 1 and phase 2, respectively (see \autoref{Fig_3}). Likewise, additional profiles for $t=0.055$~s and  $t=0.0583$~s are shown, these corresponding to the negative peak of phase 1 (third pane from the left), and the positive peak of phase 3 (right pane).}
\end{figure*}

\end{document}